\documentclass[conference]{IEEEtran}

\usepackage{amsmath,graphicx}
\usepackage{url}
\usepackage[hidelinks]{hyperref}
\usepackage{epsfig}
\usepackage{subcaption}
\usepackage{amssymb}
\usepackage{color}
\usepackage{xspace}
\usepackage{multirow} 
\usepackage{tabularx,booktabs}
\usepackage{arydshln}
\usepackage{tabu}
\usepackage{ctable}
\usepackage{enumerate,multicol}
\usepackage{multirow}
\usepackage{multicol}
\usepackage{enumitem}
\usepackage{gensymb}
\usepackage{float}
\usepackage{caption}
\usepackage{subcaption}
\usepackage{balance}
\usepackage{xspace}
\usepackage{enumerate,multicol}
\usepackage[ruled,vlined]{algorithm2e}
\usepackage{booktabs}
\usepackage[accsupp]{axessibility}  
\usepackage{subcaption}
\usepackage{titlesec}
\titlespacing*{\section}{0pt}{12pt}{12pt}  

\usepackage[
    top=0.71in,
    bottom=0.71in,
    left=0.73in,
    right=0.73in
]{geometry}

\title{Is there a relationship between Mean Opinion Score (MOS) and Just Noticeable Difference (JND)?}

\author{
Jingwen Zhu$^{1}$,
Hadi Amirpour$^{2}$,
Wei Zhou$^{3}$, 
Patrick Le Callet$^{1}$\\
$^{1}$ Nantes Université, Ecole Centrale Nantes, CNRS, LS2N, UMR 6004, Nantes, France \\
$^{2}$ Department of Information Technology (ITEC), University of Klagenfurt, Austria \\
$^{3}$ School of Computer Science and Informatics, Cardiff University, Cardiff, UK
}

\begin{document}
\maketitle
\begin{abstract}
Evaluating perceived video quality is essential for ensuring high Quality of Experience (QoE) in modern streaming applications. While existing subjective datasets and Video Quality Metrics (VQMs) cover a broad quality range, many practical use cases—especially for premium users—focus on high-quality scenarios requiring finer granularity. Just Noticeable Difference (JND) has emerged as a key concept for modeling perceptual thresholds in these high-end regions and plays an important role in perceptual bitrate ladder construction. However, the relationship between JND and the more widely used Mean Opinion Score (MOS) remains unclear. In this paper, we conduct a Degradation Category Rating (DCR) subjective study based on an existing JND dataset to examine how MOS corresponds to the 75\% Satisfied User Ratio (SUR) points of the 1\textsuperscript{st} and 2\textsuperscript{nd} JNDs. We find that while MOS values at JND points generally align with theoretical expectations (e.g., 4.75 for the 75\% SUR of the 1\textsuperscript{st} JND), the reverse mapping—from MOS to JND—is ambiguous due to overlapping confidence intervals across PVS indices. Statistical significance analysis further shows that DCR studies with limited participants may not detect meaningful differences between reference and JND videos.

\end{abstract}

\begin{IEEEkeywords}
Just Noticeable Difference (JND), Satisfied User Ratio (SUR), Mean Opinion Score (MOS), Video quality
\end{IEEEkeywords}

\section{Introduction}

Evaluating human perceived video quality is important for the streaming industry to ensure a good Quality of Experience (QoE) for end-users~\cite{zhou_perceptual_2025}. Extensive research has been conducted to collect subjective datasets for video quality assessment~\cite{wang2019youtube, sinno2018large, konvid1k, ying2020live}, and advanced objective Video Quality Metrics (VQM)~\cite{vmaf, rao2020p1204, wu2022fastquality, cover2024cpvrws,amirpour_real-time_2024,amirpour_vqm4has_2025} have been developed using learning-based methods trained on these datasets. These metrics aim to estimate human perception of quality in a computationally efficient and scalable manner, and are widely used in encoding optimization and adaptive streaming systems.

Subjective tests, in which human viewers rate the quality of videos under controlled conditions, provide the ground truth against which objective metrics are evaluated. The effectiveness of a VQM is typically measured by how well its predictions correlate with subjective scores, using statistical measures such as Pearson or Spearman correlation coefficients. High-performing methods aim to achieve strong correlation with human opinion scores, thereby ensuring their reliability in predicting user-perceived quality. As such, the design and benchmarking of objective VQMs remain closely tied to the availability and quality of subjective datasets, which capture nuanced perceptual effects that are difficult to model directly.


Most existing datasets are designed to cover the \textbf{full range} of quality levels. As shown in Table~\ref{tab:rating_scale}, a typical question in a subjective video quality study asks participants to rate the video quality using scales ranging from ``Excellent" to ``Bad" following the ACR-HR (Absolute Category Rating–Hidden Reference) method for single stimuli, or from ``Imperceptible" to ``Very annoying" according to the DCR (Degradation Category Rating) method for paired stimuli, as recommended by ITU~\cite{siti_itu_ref}. The resulting MOS or DMOS (Differential Mean Opinion Score) values serve as ground truth for training and evaluating objective video quality metrics, optimizing encoding pipelines, and more.

\begin{table}[htb]
\centering
\caption{Rating scales used in ACR for single stimuli and DCR for paired stimuli}
\begin{tabular}{c|c|c}
\toprule
Score & ACR       & DCR                            \\ \hline
5     & Excellent & Imperceptible                  \\ \hline
4     & Good      & Perceptible   but not annoying \\ \hline
3     & Fair      & Slightly   annoying            \\ \hline
2     & Poor      & Annoying                       \\ \hline
1     & Bad       & Very annoying                  \\ \bottomrule
\end{tabular}
\label{tab:rating_scale}
\end{table}

For end-users, especially premium subscribers, the expectation is to enjoy the best possible visual quality on their devices. These users are primarily concerned with whether the delivered video is perceptually indistinguishable from the original, rather than with degradations at the mid or low end of the quality spectrum. Traditional subjective quality assessment methods such as ACR and DCR evaluate across the entire quality range (Table~\ref{tab:rating_scale}), from ``Bad" to ``Excellent" or from ``Very annoying" to ``Imperceptible." While effective for general-purpose benchmarking, these methods often lack the granularity required to make fine distinctions among high-quality representations. As a result, multiple high-quality versions may receive similarly high MOS scores (\textit{e.g.}, between 4.5 and 5.0), offering limited guidance when optimizing for perceptual transparency or efficient bitrate usage at the top end of the quality ladder.

To address this limitation, the concept of Just Noticeable Difference (JND) has gained traction in recent research~\cite{zhu2024high, wang_videoset_2017, zhang_deep_2022, amirpour2024exploring}. JND defines the perceptual threshold at which a viewer just begins to notice a difference compared to an anchor—typically the pristine or reference video for the 1\textsuperscript{st} JND~\cite{lin_experimental_2015,yuan_visual_2019}. This is particularly important for applications such as bitrate ladder construction in adaptive streaming systems, where it is critical to determine the point at which increasing the bitrate no longer yields a perceptual benefit~\cite{menon_perceptually-aware_2022,takeuchi_perceptual_2018}. By identifying the boundary between perceptually lossless and lossy representations, JND allows providers to reduce the bitrate of the highest-quality representation without sacrificing visual quality. This not only improves compression efficiency but also reduces bandwidth usage and CDN costs. In this context, JND complements MOS by providing finer resolution around high-quality regions, enabling better-informed decisions in encoding, quality monitoring, and user experience optimization.

\begin{figure}[!t]
    \centering
    \includegraphics[width=0.75\linewidth]{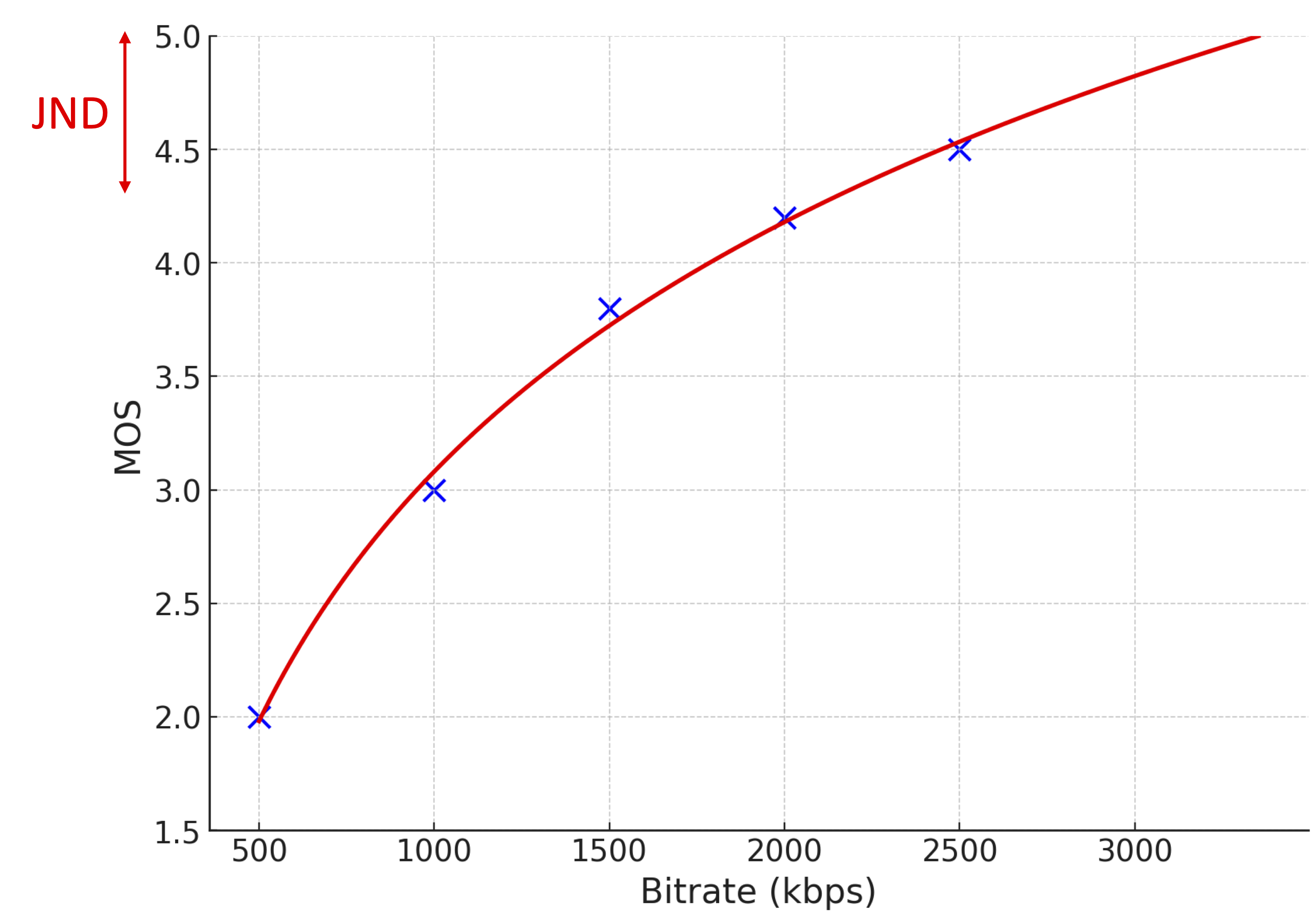}
    \caption{The relationship between MOS and JND.}
    \label{fig:mos_jnd}
    \vspace{-1em}
\end{figure}

One research question we aim to explore in this paper is the relationship between MOS and JND. Specifically, what is the typical MOS value for videos with 1\textsuperscript{st} or 2\textsuperscript{nd} JND value qualities? Should it be 5, 4.5, or another value? (see Fig.~\ref{fig:mos_jnd}) To the best of our knowledge, no prior study has explicitly examined this relationship. To address this gap, we design a subjective study based on an existing JND video dataset~\cite{zhu2022subjective}, with the goal of better understanding how MOS and JND relate to each other. Defining a relationship between the two is an open challenge, as it is unclear what MOS score corresponds to a video that sits exactly at the perceptual threshold defined by the 1\textsuperscript{st} JND.

\section{Subjective study design overview}

This section provides an overview of the subjective study design used in this work. We begin by describing the HD-VJND dataset, which forms the foundation of our analysis. We then introduce the concept of the Satisfied User Ratio (SUR), which aggregates individual JND points across observers to yield a group-level understanding of perceptual thresholds. Finally, we present our DCR-based subjective dataset, constructed using selected video samples from the HD-VJND dataset, to investigate the relationship between JND and MOS through pairwise comparison with reference content, since the original dataset lacks MOS scores.
\subsection{HD-VJND dataset}
\label{sec:hd-vjnd}
We designed the subjective study based on an existing JND dataset, namely the HD-VJND dataset~\cite{zhu2022subjective}. HD-VJND contains 180 source (SRC) videos, selected from a pool of over 600 using content selection techniques aimed at ensuring diverse visual content~\cite{ling2020towards}. Each SRC is approximately 10 seconds long and has a resolution of 1080p. All SRCs were encoded at two resolutions, 1080p and 720p, using CRF values ranging from 17 to 31 in increments of 0.25. For the JND search methodology, Robust Binary Search (RBS)~\cite{wang_videoset_2017} was employed, along with a pre-processing step to reduce search time. 

The test environment was the participants' home settings, using a selection of 55-inch displays as listed below:
\begin{itemize}
    \item SONY KD-55XH8094 x4
    \item SAMSUNG QE55Q74TATXXC x2
    \item SAMSUNG UE55TU8075U x2
    \item LG nanocell 55NAN091 x2
\end{itemize}
A total of 20 participants took part in the study, with 2 participants from each of 10 households. The participants were instructed to sit at a distance of three times the screen height, following ITU recommendations~\cite{siti_itu_ref}. They were also asked to close the curtains and turn off any lights that directly face the screen to avoid reflections. For more details, please refer to~\cite{zhu2022subjective}.

\subsection{Modeling JND Distributions with SUR}
\begin{figure}[!t]
    \centering
    \includegraphics[width=0.75\columnwidth]{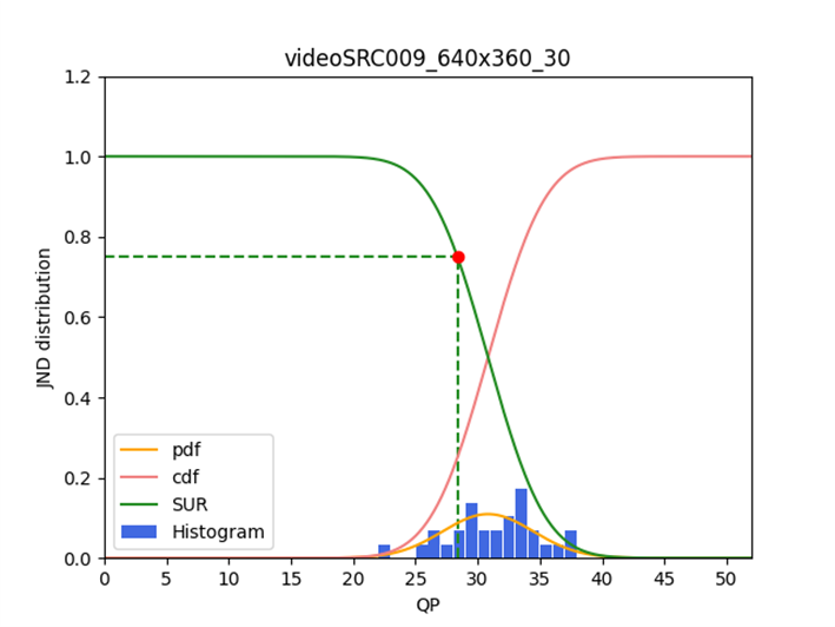}
    \caption{Illustration of the Satisfied User Ratio (SUR) for JND.}
    \label{fig:sur}
    \vspace{-1em}
\end{figure}

JND values vary across different observers. To aggregate the JND data from a group of participants, the Satisfied User Ratio (SUR) was proposed in~\cite{wang_videoset_2017}. As shown in Fig.~\ref{fig:sur}, the distribution of individual JNDs can be modeled using a probability density function (pdf) and its corresponding cumulative density function (cdf). The SUR curve is defined as $SUR=1 - cdf$, representing the proportion of users who are satisfied (\textit{i.e.,} do not notice any difference). A SUR threshold of 75\% at the 1\textsuperscript{st} JND is commonly used in prior works~\cite{wang_videoset_2017, zhu2022subjective} as an anchor for identifying the 2\textsuperscript{nd} JND, and similarly, the 75\% SUR point of the 2\textsuperscript{nd} JND is often used as the anchor for the 3\textsuperscript{rd} JND.

In the HD-VJND dataset, for smooth playback, a near-lossless compressed version of the source video (CRF=1) at the native resolution was used as the anchor for the 1\textsuperscript{st} JND search at 1080p, instead of the raw video. The 75\% SUR point of the 1\textsuperscript{st} JND is then used as the anchor for the 2\textsuperscript{nd} JND search. While the 1\textsuperscript{st} JND search is conducted solely at the encoding resolution of 1080p, the 2\textsuperscript{nd} JND search is performed twice--once at 1080p and once at 720p. This design is based on the convex hull assumption: the 1\textsuperscript{st} JND should occur at the same resolution as the source, whereas the 2\textsuperscript{nd} JND may occur at either 1080p or 720p. Conducting the 2\textsuperscript{nd} JND search at both encoding resolutions enables a better understanding of which resolution should be selected when constructing the bitrate ladder.

\subsection{DCR-Based MOS Collection}
To explore the relationship between JND and MOS, we used DCR as the subjective study protocol because, during the subjective study of JND, a reference is available for observers, and they are asked to answer whether they can perceive a difference between the distorted version and the reference. Therefore, we decided to use DCR instead of ACR because we would like to evaluate the distortion in comparison to the reference.

The DCR dataset is built based on the HD-VJND dataset. We kept the same SRCs from the HD-VJND dataset, and for each SRC, we selected 9 Processed Video Sequences (PVS), as shown in Table~\ref{tab:pvs_selection}. The principle of PVS selection is to first include the 75\% SUR points of the 1\textsuperscript{st} JND (PVS index 1) and the 2\textsuperscript{nd} JND in both 1080p and 720p (PVS indices 2 and 3). We then sampled additional video points around PVS indices 1, 2, and 3. These PVSs will help us understand the MOS at the 75\% SUR of the 1\textsuperscript{st} and 2\textsuperscript{nd} JND.

\begin{table}[htb]
\centering
\caption{PVS selection for DCR dataset}
\begin{tabular}{c|c}
\toprule
PVS index & description                      \\ \midrule
0         & Reference                        \\
1         & 75\%SUR of 1\textsuperscript{st} JND 1080p            \\
2         & 75\%SUR of 2\textsuperscript{nd} JND in 1080p         \\
3         & 75\%SUR of 2\textsuperscript{nd} JND in 720p          \\
4$\sim$7  & 4 points between PVS 1 and PVS 2 \\
8         & CRF-1 of PVS 1                   \\
9         & CRF+1 of PVS 3                   \\ \bottomrule
\end{tabular}
\label{tab:pvs_selection}
\vspace{-1em}
\end{table}

The subjective study environment was kept the same as that of the HD-VJND dataset, as described in Section~\ref{sec:hd-vjnd}. Observers were limited to a maximum of 1 hour of testing to avoid fatigue.

\section{Results and analysis}
\subsection{Data screening}
\begin{figure}[htb]
    \centering
    \begin{subfigure}[b]{0.5\textwidth}
        \centering
        \includegraphics[width=0.99\textwidth]{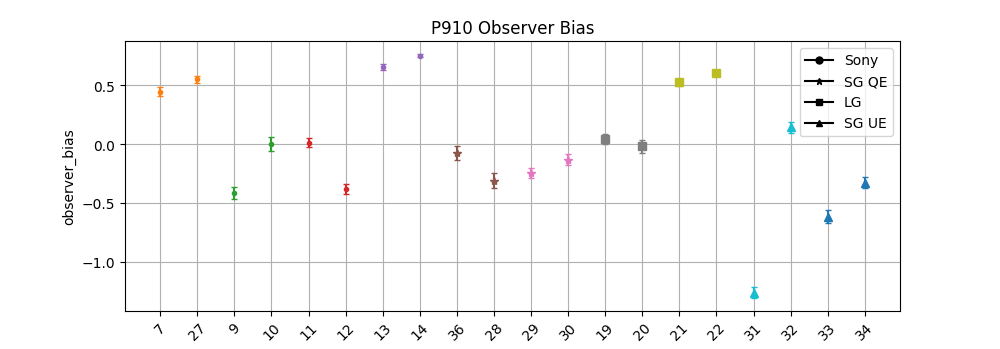}
        \caption{Bias of the 20 participants}
        \vspace{-1em}
        \label{fig:bias}
    \end{subfigure}
    \vskip\baselineskip
    \begin{subfigure}[b]{0.5\textwidth}
        \centering
        \includegraphics[width=0.99\textwidth]{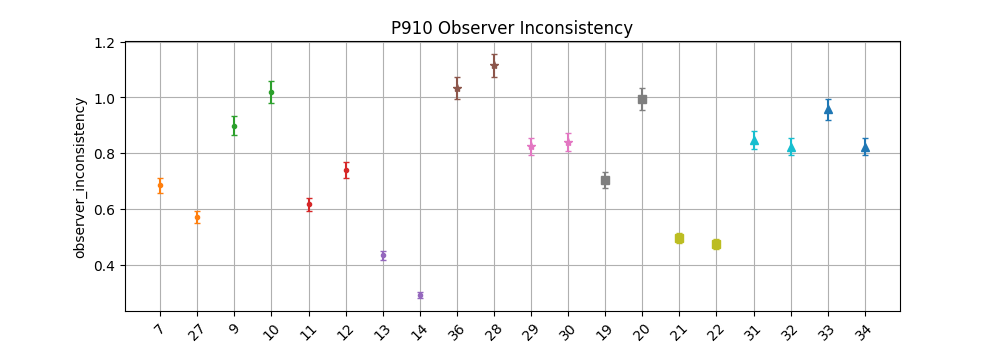}
        \caption{Inconsistency of the 20 participants}
        \label{fig:inconsistency}
    \end{subfigure}
    \caption{Bias and inconsistency of the 20 participants in the DCR dataset. Participants sharing the same home environment are indicated by the same color, while the same marker shape denotes the same TV reference.}
    \vspace{-1.5em}
    \label{fig:bias_inconsistancy}
\end{figure}

After collecting the DCR dataset, we first perform data screening following ITU-T Recommendation P.910~\cite{itu2022BT910}. The bias and consistency of the 20 participants are shown in Fig.~\ref{fig:bias_inconsistancy}. 

Observers are grouped by home environment, with the same color indicating participants from the same household. For example, Observer 7 and Observer 27 share the same home environment and use the same Sony TV. An interesting observation is that participants from the same home environment generally exhibit similar bias and inconsistency. However, this pattern is not observed for participants using the same TV model in different environments, suggesting that viewing conditions may have a greater impact than display type. This effect might also be influenced by the fact that cohabiting participants are often couples who share similar educational backgrounds and habits, potentially leading to comparable behavior during the subjective study.

\subsection{MOS at JND Anchor Points}
We compute the MOS scores of the PVS index 0,1,2,3 across the entire collected DCR dataset; the results are shown in Table~\ref{tab:mos_jnd}. The raw scores $S_{pvs}$ are mapped to $S'_{pvs}$  using Eq.~(\ref{eq:mapping_os}) so that the MOS of the reference videos is 5.

\vspace{-1em}
\begin{equation}
\label{eq:mapping_os}
    {S'_{pvs}} = {S_{pvs}} - {S_{ref}} + 5
\end{equation}
\vspace{-1em}

\begin{table}[htb]
\centering
\caption{Average of MOS and 95\% Confidence Intervals (CI) at PVS index 0–3 across the entire dataset}
\begin{tabular}{c|c|c|c}
\toprule
\begin{tabular}[c]{@{}c@{}}PVS \\ index\end{tabular} & description & Average of   MOS & 95\%CI \\ \midrule
0 & Reference                  & 5    & {[}4.90, 5.10{]} \\
1 & 75\%SUR of 1\textsuperscript{st}  JND 1080p  & 4.79 & {[}4.64, 4.93{]} \\
2 & 75\%SUR of 2\textsuperscript{nd}   JND 1080p & 4.51 & {[}4.31, 4.72{]} \\
3 & 75\%SUR of 2\textsuperscript{nd}   JND 720p  & 4.27 & {[}4.01, 4.52{]} \\ \bottomrule
\end{tabular}
\label{tab:mos_jnd}
\end{table}

Ideally, for the video that corresponds to the 75\% SUR of the 1\textsuperscript{st} JND (\textit{i.e.,} PVS index 1), the definition of SUR implies that 75\% of observers cannot perceive any difference between this video and the reference. As a result, during a DCR study, these observers would vote ``Imperceptible" with a score of 5 (see Table~\ref{tab:rating_scale}). The remaining 25\% of observers, who do notice a difference, would vote ``Perceptible but not annoying" with a score of 4. Therefore, the ideal MOS for the 75\% SUR point of the 1\textsuperscript{st} JND should be computed using Eq.~(\ref{eq:ideal_mos}), resulting in a value of 4.75. 
\begin{equation}
    MOS{_{ideal}}(75\%\ \text{SUR}_{\textbf{\text{1st JND}}}) = 5 \times 0.75 + 4 \times 0.25 = 4.75
    \label{eq:ideal_mos}
\end{equation}

For the 75\% SUR of the 2\textsuperscript{nd} JND, the video with PVS index 1 was used as the anchor during the JND search. Ideally, if we use this same video (PVS index 1) as the reference in the DCR subjective study instead of PVS index 0, we should obtain an MOS of 4.75 for the 75\% SUR of the 2\textsuperscript{nd} JND. However, in the actual DCR study, we used PVS index 0 as the reference (which was also the anchor for the 1\textsuperscript{st} JND search). Since this reference is of higher quality, we would expect to obtain a lower MOS than 4.75 due to the increased perceptual gap. The ideal MOS for the 75\% SUR point of the 2\textsuperscript{nd} JND (\textit{i.e.,} PVS index 2) is computed using Eq.~(\ref{eq:delta}) and Eq.~(\ref{eq:2ndjnd}).

\begin{equation}
    \Delta  = 5 - MO{S_{ideal}}(75\%\ \text{SUR}_{\text{1st JND}}) = 0.25
    \label{eq:delta}
    \vspace{-1em}
\end{equation}

{
\small
\begin{align}
    MOS_{\text{ideal}}(75\%\ \text{SUR}_{\textbf{2nd JND}}) 
    &= MOS_{\text{ideal}}(75\%\ \text{SUR}_{\textbf{1st JND}}) - \Delta \notag \\
    &= 4.75 - 0.25 = 4.5
    \label{eq:2ndjnd}
\end{align}

}

It can be observed from Table~\ref{tab:mos_jnd} that the 75\% SUR of the 1\textsuperscript{st} JND is 4.79, which is close to the ideal MOS for this video. The MOS of the 75\% SUR point of the 2\textsuperscript{nd} JND in 720p is slightly lower than that in 1080p. The 1080p version is closer to the ideal MOS computed using Eq.~(\ref{eq:delta}) and Eq.~(\ref{eq:2ndjnd}).

\begin{figure}[htb]
    \centering
    \includegraphics[width=0.8\linewidth]{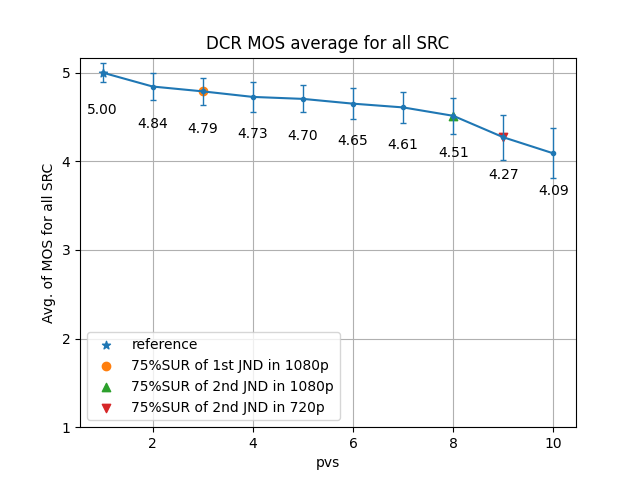}
    \caption{DCR MOS value for 75\%SUR in 1\textsuperscript{st} JND and 2\textsuperscript{nd} JND in 1080p and 2\textsuperscript{nd} JND in 720p. (Note: the original MOS are mapped to make the reference MOS = 5)}
    \vspace{-1em}
    \label{fig:mos_avg}
\end{figure}

Fig.~\ref{fig:mos_avg} presents the mean DCR MOS across the entire dataset for all PVS indices, along with their 95\% CIs. As expected, the MOS generally decreases as compression strength increases. The 95\% CI of the average MOS at the 75\% SUR point of the 1\textsuperscript{st} JND is $[4.64, 4.93]$, which can be interpreted as a 95\% probability that the true MOS for this JND level falls within this range.

Although the mean MOS values differ slightly across PVS indices, their 95\% CIs often overlap. As a result, the relationship between JND and MOS is inherently one-directional:

\begin{itemize}
\item \textbf{JND to MOS is feasible:} For example, we can reasonably infer that the MOS corresponding to the 75\% SUR of the 1\textsuperscript{st} JND lies within 95\% CI $[4.64, 4.93]$.
\item \textbf{MOS to JND is ambiguous:} For instance, if a video receives a MOS of 4.7 in a subjective study, we cannot confidently determine whether it corresponds to the 75\% SUR of the 1\textsuperscript{st} JND or the 2\textsuperscript{nd} JND.
\end{itemize}

Previous works~\cite{amirpour_between_2022, zhu2023enhancing} have investigated the mapping relationship between video quality metrics (VQMs), such as VMAF~\cite{vmaf}, and the 1\textsuperscript{st} JND. Since most VQMs—including VMAF—are trained on MOS or DMOS values obtained from subjective studies, the mapping from VQM to JND also exhibits a one-way nature. In other words, while JND levels can be associated with specific VQM scores, the reverse mapping is not uniquely defined—an observation also noted in~\cite{zhu2024beyond}.

\subsection{Statistical Significance Analysis}
In the previous section, we obtained the mean DCR MOS for different JND orders. In this section, for each SRC, we conduct a one-way ANOVA significance test across different PVS indices, and compute the percentage of SRCs that show statistically significant differences out of the total number of SRCs. The results are summarized in Table~\ref{tab:anova}. It can be observed that only 10.71\% of the SRCs show a statistically significant difference (p-value $<$ 0.05) between the reference and the 75\% SUR point of the 1\textsuperscript{st} JND. This indicates that the DCR subjective study is generally unable to conclude a significant difference between PVS 0 (reference) and PVS 1 (75\% SUR of the 1\textsuperscript{st} JND), likely because the quality of the two videos is too similar and the number of participants is limited. However, the MOS values still reflect a slight difference between these two videos, as shown in Table~\ref{tab:mos_jnd} and Fig.~\ref{fig:mos_avg}.

\begin{table}[htb]
\centering
\caption{Percentage of significantly different videos on entire datasets with one-way ANOVA}
\vspace{-0.75em}
\begin{tabular}{c|c}
\toprule
PVS index  pair & significant  different percentage \\ \midrule
(0, 1)          & 10.71\%                           \\
(0, 2)          & 38.40\%                           \\
(0, 3)          & 64.29\%                           \\
(2, 3)          & 6.25\%                            \\
(1, 2)          & 7.14\%                            \\
(1, 3)          & 30.36\%                           \\ \bottomrule
\end{tabular}
\label{tab:anova}
\vspace{-2em}
\end{table}

\section{Conclusion}
In this paper, we explored the relationship between JND and MOS obtained from a DCR subjective study, aiming to answer: what are the MOS values corresponding to the 1\textsuperscript{st} and 2\textsuperscript{nd} JNDs? We first designed and conducted a DCR study based on the existing HD-VJND dataset. By analyzing participant behavior across different environments and displays, we found that participants in the same setting tend to show similar bias and inconsistency.

Our results revealed a one-way mapping between JND and DCR MOS. Specifically, while we can reasonably estimate the MOS corresponding to a given JND level (\textit{e.g.}, the 75\% SUR of the 1\textsuperscript{st} JND), the reverse does not hold: knowing the MOS of a video does not indicate its JND level.

We also performed statistical significance analysis, which showed that a DCR study with few participants may fail to detect meaningful differences between JND videos and the reference, as their quality is often too close. In future work, we plan to increase the number of participants to improve the discriminative power of the DCR protocol.

\balance

\bibliographystyle{IEEEbib}
\bibliography{refs,ref_hadi}

\end{document}